\documentclass[twocolumn,superscriptaddress,aps,preprintnumbers,amsmath,amssymb,,sort&compress,nofootinbib,
]{revtex4-1}
\usepackage{amsfonts,amsmath,amssymb,ascmac,bm,tensor}
\usepackage{comment}
\usepackage{slashed}
\usepackage{color}
\usepackage[mathscr]{eucal}
\usepackage[utopia]{mathdesign}
\usepackage[utf8]{inputenc}

\usepackage{cancel}

  \usepackage{graphicx}     %   usepackage without driver option
  \usepackage[bookmarksopen,colorlinks=true,linkcolor=bblue,citecolor=bblue,urlcolor=ppink]{hyperref}
\definecolor{red}{rgb}{1,0,0}
\definecolor{darkred}{rgb}{0.6,0,0}
\definecolor{darkgreen}{rgb}{0.992447,0.623778,0.034597}
\definecolor{ppink}{rgb}{1,0.4,0.4}
\definecolor{bblue}{rgb}{0.284602,0.317763,0.963947}
\newcommand{\prn}[1]{\left( {#1} \right)}
\newcommand{\com}[1]{\left[ {#1} \right]}

\newcommand{\dd}{\mathrm{d}}
\newcommand{\Mpl}{M_{\rm Pl}}

\def\Mpl{M_{\rm Pl}}

\newcommand{\dif}[2]{\frac{\mathrm{d} #1}{\mathrm{d} #2}}

\newcommand{\calP}{\mathcal{P}}
\newcommand{\ns}{n_{{}_\mathrm{S}}}

\makeatletter
\newcommand\footnoteref[1]{\protected@xdef\@thefnmark{\ref{#1}}\@footnotemark}
\makeatother

\begin{document}

%%%%%%%%%%%%%%%%%%%%%%%%%%%
%%%%%%%%%%% Title %%%%%%%%%%%
%%%%%%%%%%%%%%%%%%%%%%%%%%%

%%paper
\title{
Inflationary Primordial Black Holes as All Dark Matter
}
\author{Keisuke Inomata}
\affiliation{ICRR, University of Tokyo, Kashiwa, 277-8582, Japan}
\affiliation{Kavli IPMU (WPI), UTIAS, University of Tokyo, Kashiwa, 277-8583, Japan}
\author{Masahiro Kawasaki}
\affiliation{ICRR, University of Tokyo, Kashiwa, 277-8582, Japan}
\affiliation{Kavli IPMU (WPI), UTIAS, University of Tokyo, Kashiwa, 277-8583, Japan}
\author{Kyohei Mukaida}
\affiliation{Kavli IPMU (WPI), UTIAS, University of Tokyo, Kashiwa, 277-8583, Japan}
\author{Yuichiro Tada}
\affiliation{ICRR, University of Tokyo, Kashiwa, 277-8582, Japan}
\affiliation{Kavli IPMU (WPI), UTIAS, University of Tokyo, Kashiwa, 277-8583, Japan}
\author{Tsutomu T.~Yanagida}
\affiliation{Kavli IPMU (WPI), UTIAS, University of Tokyo, Kashiwa, 277-8583, Japan}

\begin{abstract}
\noindent
Following a new microlensing constraint on 
primordial black holes (PBHs) with $\sim10^{20}$--$10^{28}$\,g~\cite{takada},
we revisit the idea of PBH as all Dark Matter (DM).
We have shown that
the updated observational constraints suggest the viable mass function 
for PBHs as all DM to have a peak at $\simeq 10^{20}$\,g with a small width $\sigma \lesssim 0.1$,
by imposing observational constraints on an extended mass function in a proper way.
We have also provided an inflation model that successfully generates PBHs as all DM
fulfilling this requirement.

\end{abstract}

\date{\today}
\maketitle
\preprint{IPMU17-0009}

%%%%%%%%%%%%%%%%%%%%%%%%%%%%%%%%%
%%%%%%%%%%% Introduction %%%%%%%%%%%
%%%%%%%%%%%%%%%%%%%%%%%%%%%%%%%%%
\section{Introduction}\label{sec: intro}

Dark matter (DM) is one of the outstanding problems,
which motivates us to seek for new particle physics models beyond the Standard Model (SM).
Its existence is well established by the astrophysical and cosmological observations.
However, we still do not know most of its property.
For instance, its possible mass scale ranges large order of magnitude 
from $10^{-31}$\,GeV to $10^{50}$\,GeV. 
Primordial black hole (PBH)~\cite{Hawking:1971ei,Carr:1974nx,Carr:1975qj} 
resides in the heavy end of various candidates of DM.
It behaves as cold matter, is stable for sufficiently heavy ones, and thus is a perfect candidate of DM.
In particular, we do not need new particles beyond the SM.
Therefore, whether or not PBHs can be a dominant component of DM is an important issue for particle physics.

PBHs are formed if large density perturbations collapse overcoming the pressure forces. Cosmic inflation can be a source of such large density perturbations. For instance, if the potential has a plateau during the inflationary epoch, large superhorizon fluctuations are produced, and they eventually collapse to form PBHs at the horizon reentry. Although the amplitude of scalar perturbations is strictly constrained at the large scale, say $0.0002$--$1$\,Mpc$^{-1}$~\cite{Ade:2015xua}, by the cosmic microwave background (CMB) observation, they can be significant at the smaller scale, which opens up a possibility to produce a sizable amount of PBHs comparable to the current DM density.

Recently, making use of the Subaru Hyper Suprime-Cam (HSC) dense cadence data, Niikura et al.~\cite{takada} put the most stringent upper bounds on the PBH abundance in the mass range $10^{20}\text{--}10^{28}\,\mathrm{g}$.\footnote{
\label{ns}
	Throughout this paper, we assume a conservative value for the abundance of DM inside the globular clusters,
	and neglect the NS constraints~\cite{Kusenko:2013saa,Carr:2016drx,Clesse:2016vqa}. 
}
In this paper, we show that when we combine it with other observational constraints, only the mass region $\simeq 10^{20}\,\mathrm{g}$ remains for PBHs as all DM,
by imposing observational constraints on the extended mass function in a proper way.
We also show that the PBHs, which have an inflationary origin, can still be a dominant component of DM of mass $\simeq10^{20}\,\mathrm{g}$, taking the model~\cite{Kawasaki:1997ju,Kawasaki:1998vx,Frampton:2010sw,Kawasaki:2012kn, Kawasaki:2016ijp,Kawasaki:2016pql,Inomata:2016rbd} as an example.

%%%%%%%%%%%%%%%%%%%%%%%%%%%%%%%%%%%%
\section{Formation of PBHs during Inflation}
\label{sec:pbh_formation}
%%%%%%%%%%%%%%%%%%%%%%%%%%%%%%%%%%%%

The property of PBH is characterized by its mass and abundance.
In the following, we briefly summarize the formation of PBHs by large superhorizon fluctuations.
We adopt the conventional analysis for the formation of PBHs~\cite{Carr:1975qj,Green:1997sz}.
See \cite{Niemeyer:1999ak,Shibata:1999zs,Musco:2004ak,Musco:2012au,Harada:2013epa,Nakama:2013ica} for attempts to refine the simple analysis.

When an over-dense region above the threshold, $\delta \rho / \rho > \delta_c$,  reenters the horizon, 
with the threshold being $\delta_c$,
it may overcome the pressure and collapse to form the PBHs.
There exist many attempts to pin down the threshold value,
but here we adopt the conventional one $\delta_c = 1/3$ as a reference.
In the simple analysis,
the mass of PBH is proportional to the horizon mass at that time.
Thus, it is estimated by
\begin{align}
	M(k)  
	&= \left. \gamma \rho \frac{4 \pi H^{-3}}{3} \right|_{k = aH}
	\simeq \frac{\gamma M_\text{eq}}{\sqrt{2}}
	\prn{ \frac{g_{\ast,\text{eq}}}{g_\ast} }^\frac{1}{6}
	\prn{ \frac{k_\text{eq}}{k} }^2 \\[.5em]
	&\simeq 10^{20}\,\mathrm{g}\,
    \left(
    \frac{\gamma}{0.2}
    \right)
    \left(
    \frac{g_\ast}{106.75}
    \right)^{- \frac{1}{6}}
    \left(
    \frac{k}{7\times 10^{12} \,\textrm{Mpc}^{-1}}
    \right)^{-2}.
	\label{eq:pbhmass}
\end{align}
Here we used an approximation that the effective degrees of freedom for energy density $g_*$ is almost equal to that for entropy density $g_{*s}$.
$\gamma$ represents the ratio between the  PBH mass and the horizon mass,
which is estimated as $\gamma \simeq 3^{-3/2}$ in the simple analytical result~\cite{Carr:1975qj}.
$M(k)$ denotes the mass of PBH that is formed 
when the comoving momentum $k$ reenters the horizon. $k_\text{eq}$ is the comoving momentum at the matter-radiation equality, \textit{i.e.,} 
$k_\text{eq} = a_\text{eq} H_\text{eq}$.
$M_\text{eq}$ is the horizon mass at the same time.

The formation rate of PBHs with mass $M$, $\beta (M)$, is given by the probability of exceeding the threshold $\delta_c$.
We assume that the density perturbation is governed by the Gaussian statistics.
Then, the formation rate is given by
\begin{align}
	\beta (M) = \int_{\delta_c}
	\frac{\dd \delta}{\sqrt{2 \pi \sigma^2 (M)}} \, e^{- \frac{\delta^2}{2 \sigma^2 (M)}}
	\simeq 
	\frac{1}{\sqrt{2 \pi}} \frac{1}{\delta_c / \sigma (M)} \, e^{- \frac{\delta_c^2}{2 \sigma^2 (M)}}.
\end{align}
$\sigma^2 (M)$ represents the standard deviation of the coarse-grained density contrast for the PBH mass of $M$~\cite{Young:2014ana}
\begin{align}
	\sigma^2 (M(k)) = \int \dd \ln q W^2 (q k^{-1}) \frac{16}{81} \prn{q k^{-1}}^4
	\mathcal P_\zeta (q),
\end{align}
where $W$ is the Fourier transform of the window function smearing over $k^{-1}$,
and we adopt the Gaussian window $W (x) = e^{ - x^2 / 2}$.
At the horizon reentry, a fraction of the total energy of the Universe,
$\gamma \beta (M(k)) \rho |_{k = aH}$, turns into PBHs.
After their formation, 
$\rho_\text{PBH}/\rho$ grows inversely proportional to the cosmic temperature until the matter-radiation equality, since PBHs behave as matter.
Thus, the abundance of PBHs with mass $M$ over logarithmic mass interval $\dd \ln M$ may be estimated as
\begin{align}
	&f_\text{PBH} (M) \equiv \frac{\Omega_\text{PBH}(M)}{\Omega_c }
	= \left. \frac{\rho_\text{PBH}}{\rho_m} \right|_\text{eq} \frac{\Omega_mh^2}{\Omega_ch^2}
	= \prn{\frac{T_M}{T_\text{eq}} \frac{\Omega_mh^2}{\Omega_ch^2}} \gamma \beta (M) \nonumber\\
	&\simeq
	\prn{ \frac{\beta (M)}{8.0 \times 10^{-15}} }
	\prn{\frac{0.12}{\Omega_ch^2}}
	\prn{\frac{\gamma}{0.2}}^\frac{3}{2}
	\prn{ \frac{106.75}{g_{\ast} (T_M)} }^\frac{1}{4}
	\prn{ \frac{M}{10^{20}\,\mathrm{g}} }^{-\frac{1}{2}}.
\end{align}
$T_\text{eq}$ indicates the temperature at the matter-radiation equality and
$T_M$ is the temperature at the formation of a PBH with a mass $M$.
$\Omega_m$ ($\Omega_c$) is the current density function of matter (DM) where we used the recent Planck's result $\Omega_ch^2\simeq0.12$~\cite{Ade:2015xua}.
The total abundance of PBH is obtained from
\begin{align}
	\Omega_{\text{PBH,tot}} = \int \dd \ln M\, \Omega_\text{PBH} (M).
\end{align}
As one can see from obtained equations,
a sizable amount of PBHs is produced if the scalar perturbations are significant $\mathcal P_\zeta \sim 10^{-2}$.
Thus, the problem is how to generate such large scalar perturbations
without conflicting the CMB observation.

\paragraph*{\bf Double Inflation.}
To make our discussion concrete, we adopt the double inflation scenario proposed in Ref.~\cite{Kawasaki:1997ju} as an example,
which involves new inflation as the second inflation.
(See also Refs.~\cite{Kawasaki:1998vx,Frampton:2010sw,Kawasaki:2012kn,Kawasaki:2016ijp,Kawasaki:2016pql}.)
We assume chaotic inflation as the pre-inflation before new inflation.
The pre-inflation dynamically determines the initial condition of the new inflation,
which solves the crucial drawback of the new inflation, namely the initial condition problem~\cite{Izawa:1997df}.
In addition, 
the chaotic inflation is responsible for the large-scale perturbations, $k \lesssim 1$\,Mpc$^{-1}$, observed by Planck,
and hence the new inflation can be free from the COBE normalization, which allows much larger scalar perturbations at the smaller scale.

Hereafter, we phenomenologically take the following potential:\footnote{
	In general, one also expects the Planck-suppressed operators for kinetic terms. Here we suppressed them for simplicity. See Ref.~\cite{Inomata:2016rbd} for their possible effects on the formation of PBHs.
}
%(See Appendix for its possible origin)
%%
\begin{align}
	V (\phi ,\varphi) 
	=& V_\text{ch} (\phi) 
	+ V_\text{stb} (\phi ,\varphi)
	+ V_\text{new} (\varphi), \\
	V_\text{new} (\varphi)
	=&
	\prn{ v^2 - g \frac{\varphi^n}{\Mpl^{n-2}} }^2 
	-\kappa v^4\frac{\varphi^2}{2\Mpl^2}
	-\varepsilon v^4 \frac{\varphi}{\Mpl}, \\
	V_\text{stb} (\phi,\varphi) =&
	c_\text{pot} \frac{V_\text{ch} (\phi) }{2 \Mpl^2} \varphi^2. 
	\label{eq:pot}
\end{align}
%%
%%%%%%%%%%%%%%%%%%%%%%%%%%%%%%%%%

%%%%%%%%%%%%%%%%%%%%%%%%%%%%%%%%%
$\phi$ and $ \varphi$ are the inflatons responsible for chaotic inflation and the new inflation respectively.
$v$ is the scale of the new inflation.
$c_\text{pot}$, $g$, $\kappa$, and $\varepsilon$ are dimensionless parameters.
$V_\text{ch} (\phi)$ and $V_\text{new} (\varphi)$ are the potentials for chaotic inflation and 
the new inflation respectively.
$V_\text{stb} (\phi, \varphi)$ stabilizes $\varphi$ during chaotic inflation for $c_\text{pot} \gtrsim \mathcal O (1)$.
For simplicity, we assume $V_\text{ch} (\phi) \simeq m_\phi^2 \phi^2  / 2$.
A slight modification of the large field value regime to accommodate the CMB observation is straightforward.

Let us briefly sketch the dynamics of this model.
First, chaotic inflation takes place.
The field value of $\varphi$ is determined by the balance between the linear term $\varepsilon v^4 \varphi / \Mpl$
and $V_\text{stb}$ as $\varphi \sim \varepsilon v^4 \Mpl / V_\text{ch}$.
After the pre-inflation, $\phi$ starts to oscillate and keeps stabilizing $\varphi$ until the energy of chaotic inflaton becomes small
enough as $V_\text{ch}\sim v^4$ and the second new inflation starts.
Thus, the initial field value of the new inflation is roughly $\varphi_i \sim \varepsilon \Mpl$.
The new inflation continues until the slow roll condition is violated
at $\varphi_e \sim (v^2 \Mpl^{n-4} / (2n(n-1)g)  )^{1/(n-2)}$.
Typical sizes of parameters to achieve successful inflation are
$|\kappa | \lesssim \mathcal{O} (0.1)$ and $\varepsilon \ll (v^2  / (2n(n-1)g \Mpl^2)  )^{1/(n-2)}$.
After the end of the new inflation,
$\varphi$ oscillates around its potential minimum with a mass scale of $m_\varphi \sim n (v^2/\Mpl) (v^2 / g \Mpl^2)^{-1/n}$.
Assuming that the inflaton decays via a dimension-five Planck-suppressed operator,
one can estimate the reheating temperature as
$T_\text{R} \sim (90 / \pi^2 g_\ast)^{1/4} \sqrt{m_\varphi^3 / \Mpl}$.
We adopt this value of the reheating temperature in the following analysis.

Finally, let us briefly describe the power spectrum of scalar perturbations during the new inflation. The curvature perturbation during the new inflation may be evaluated by
\begin{align}
	\mathcal P_\zeta
	& = \prn{ \frac{H}{2 \pi} }^2 \prn{ \frac{H}{\dot \varphi} }^2 
	\simeq  \frac{V^3}{12 \pi^2 \Mpl^6 V'{}^2} \nonumber \\
	&\simeq \com{ \frac{v^2}{2 \sqrt{3} \pi \Mpl^2} 
	\frac{1}{\varepsilon + 
	\kappa \frac{\varphi}{\Mpl}
	+ 2n g \frac{\varphi^{n-1}}{v^2 \Mpl^{n-3}}}}^2.
\end{align}
The curvature perturbation becomes large at the beginning of the new inflation.
For $\varphi \sim \varphi_i$, the value can be estimated as 
$\mathcal P_\zeta \sim v^4 / (12 \pi^2  \Mpl^4 \varepsilon^2)$.
One can see that the scalar perturbations can be sizable for
$\varepsilon = \alpha v^2 / \Mpl^2$ with $\alpha \sim \mathcal O (1)$.
In the following discussion, we consider the case with $\kappa > 0$.\footnote{
 	For $\kappa  < 0$, the curvature perturbations can also be large at $\varphi \sim \varphi_\ast$
	if there exists the flat inflection point $V' (\varphi_\ast) \simeq V'' (\varphi_\ast) \simeq 0$.
	In this paper, we do not consider this case.
	See Ref.~\cite{Kawasaki:2016pql}.
}
After the beginning of the new inflation, the scale dependence of the power spectrum is given by, in the slow-roll limit,
\begin{align}
	\dif{\log\calP_\zeta}{\log k}=\ns-1=-6\epsilon_V+2\eta_V\simeq2\eta_V\simeq2\kappa,
\end{align}
where $\epsilon_V=( \Mpl^2 / 2)\left(V^\prime / V\right)^2$ and $\eta_V=\Mpl^2V^{\prime\prime}/V$
are the slow-roll parameters.
Therefore the power spectrum can be approximated by
\begin{align}
	\mathcal P_\zeta \sim
	\begin{cases}
		\left. \mathcal P_\zeta \right|_\text{ch} \sim 10^{-9} &\text{for}~k_i \gtrsim k, \\[.5em]
		\frac{v^4 } {12 \pi^2 \Mpl^4 \varepsilon^2}
		\prn{ \frac{k_i}{k} }^{2 \kappa} &\text{for} ~ k \gtrsim k_i,
	\end{cases}
	\label{eq:power_spec}
\end{align}
where $k_i$ is a comoving momentum which exits the horizon at the beginning of the new inflation.
One can see that a sizable $\kappa \sim \mathcal O (0.1)$ yields a sharp spectrum,
while the slow roll condition enforces $\kappa \ll 1$.\footnote{
 	Planck-suppressed corrections on kinetic terms allow a much steeper spectrum. See Ref.~\cite{Inomata:2016rbd}. 
}
Here we estimate the scalar power spectrum in the analytic slow-roll approximation, but we show the 
full numerical result in the linear order in Fig.~\ref{fig:pbh}.

%%%%%%%%%%%%%%%%%%%%%%%%%%%%%%%%%
%%%%%%% Constraints on Inflationary PBHs %%%%%%%
%%%%%%%%%%%%%%%%%%%%%%%%%%%%%%%%%
\section{Observational Constraints}\label{sec:const_pbh}

In this section, we briefly summarize observational constraints imposed on PBHs.
See Refs.~\cite{Carr:2009jm,Carr:2016drx} for a review.
We may roughly split them into two classes:
(a) constraints related to the current abundance of PBHs,
and (b) constraints involving
physical processes which might accompany the PBH formation or its time-evolution.
See also Fig.~\ref{fig:pbh} as an illustration,
but be careful that constraints shown in Fig.~\ref{fig:pbh} assume a monochromatic mass function.
Thus, we have to extend the treatment so as to put constraints on an extended mass function
as concretely discussed in Appendix.
See also discussion in the next section and Fig.~\ref{fig:full_scan}.

Let us start with the constraints of the class (a) from the lighter ones.

\paragraph*{\bf Extra-galactic gamma-ray background (green).}
First of all, PBHs lighter than $6 \times 10^{14}\, \mathrm{g}$ evaporate within the current age of the Universe,
which is out of our interest.
Though PBHs with $6 \times 10^{14} \lesssim M_\text{PBH} \lesssim 10^{17}\,\mathrm{g}$ remain by now,
they emit a sizable amount of photons that contribute to the extragalactic photon background.
EGRET and Fermi LAT put constraints on this mass range~\cite{Carr:2009jm}.
This constraint is shown by the green line with shade in Fig.~\ref{fig:pbh}.

\paragraph*{\bf Gravitational lensing (blue).}
Above $10^{17}\, \mathrm{g}$, gravitational lensing constraints come into play,
which is caused when a compact object passes through our line of sight towards known sources.
PBHs with $10^{17} \, \mathrm{g} \lesssim M_\text{PBH} \lesssim10^{19} \, \mathrm{g}$
are constrained,
since we do not see any femtolensing events from gamma-ray bursts of known redshift observed by the Fermi Gamma-ray Burst Monitor~\cite{Barnacka:2012bm}.
The other three constraints from $10^{20}\, \mathrm{g}$ to $10^{35}\, \mathrm{g}$
basically stem from null observation of micro/millilensing events.
The difference comes from the sources they used and the sensitive time scale of lensing events,
which results in different mass ranges of PBHs.
The constraint from Kepler satellite utilizes near stars $\sim 1 \,\textrm{kpc}$ slightly out of our galaxy plane~\cite{Griest:2013esa}.
The MACHO/EROS collaboration observed stars in Large and Small Magellanic Clouds, 
$\sim 50$ and $60$\,\textrm{kpc}~\cite{Tisserand:2006zx}.
Recently, a very severe constraint ranging from $10^{20}$ to $10^{28}$\,\textrm{g} 
has come out by using the Subaru HSC data~\cite{takada}.
Source stars are in Messier 31 (M31), $\sim 770$\,\textrm{kpc}.
The large distance of M31 and
its short cadence data enable us to probe PBHs of smaller masses.
These constraints are shown by blue lines with shades in Fig.~\ref{fig:pbh}.

\paragraph*{\bf Dynamical constraints (orange).}
PBHs may collide with astrophysical objects and could leave observational signatures.
Here we introduce three constraints, which are relevant in the following discussion.
The others that we will not mention here are basically less stringent than different constraints.
In Refs.~\cite{Capela:2012jz,Capela:2013yf}, it is claimed that neutron stars (NSs) in the globular clusters may capture PBHs
which destroys NSs immediately.
The existence of NSs in the globular clusters puts constraints on PBHs with $10^{16}$\,g to $10^{25}$\,g.
However, it is argued for instance in Refs.~\cite{Kusenko:2013saa,Carr:2016drx,Clesse:2016vqa} that
the amount of DM inside globular clusters can be much smaller than their assumption, and the constraints are evaded for dark matter densities below $\sim 10^2\,\textrm{GeV\,cm}^{-3}$.
Thus, we will not show the constraint from the NS-capture, for it can be avoided for conservative values of DM density.
Another constraints come from a dynamical heating of stars in the cluster or ultra-faint dwarf galaxies~\cite{Brandt:2016aco}.
It is claimed that a conservative limit comes from the entire sample of compact ultra-faint dwarf galaxies,
which may exclude PBHs with $10^{34}$--$10^{39}$\,g.
Also, the dynamical heating around the trajectory of PBHs may lead to the explosion of white dwarfs
as supernovae~\cite{Graham:2015apa}. It is argued that the shape of the observed distribution of white dwarfs rules out PBHs
with $10^{19}$--$10^{20}\,\textrm{g}$.
Ref.~\cite{Gaggero:2016dpq} claims 
that PBHs cannot constitute all the DM for $M \sim \mathcal O (10) M_\odot$
by using a new accretion constraint on PBHs at the galactic center via the radio and X-ray.
These constraints are shown by orange lines with shades in Fig.~\ref{fig:pbh}.

Then we discuss observational constraints (b).
\paragraph*{\bf Cosmic Microwave Background (red).}
Massive PBHs may accrete the gas during the recombination epoch.
The emission of radiation associated with the accretion could provide observable signatures
in CMB spectrum and anisotropies.
By using FIRAS/WMAP data, \cite{Ricotti:2007au} puts severe constraints on PBHs above $10^{33}$\,\textrm{g}.\footnote{
	See also a recent update discussed in Ref.~\cite{Chen:2016pud}.
}
However, it is claimed in Refs.~\cite{Bird:2016dcv,Clesse:2016vqa,Carr:2016drx} that the constraints require modeling
of complicated physical processes and thus the associated uncertainties are difficult to estimate.
Moreover, recent conservative analyses~\cite{Ali-Haimoud:2016mbv,Blum:2016cjs,Horowitz:2016lib} show that the bound can be much weaker,
allowing PBHs with masses of $M \lesssim \mathcal O (10) M_\odot$ -- $\mathcal O (10^2) M_\odot$.
This constraint is shown by a red line in Fig.~\ref{fig:pbh}.

\paragraph*{\bf Gravitational Waves from second order effects.}
In order to generate a sizable amount of PBHs,
large scalar perturbations during the inflation are required,
$\mathcal P_\zeta \sim 10^{-2}$, as discussed in the previous section.
While such large scalar perturbations may lead to the PBH-formation
when they reenter the horizon, they can simultaneously generate a substantial amount of GWs
as second order effects~\cite{Saito:2008jc,Saito:2009jt,Bugaev:2009zh,Bugaev:2010bb}.
This is because the energy-momentum tensor of scalar perturbations acts as the source term
in the equation of motion for GWs.
Since GWs are produced when the scalar perturbations reenter the horizon,
the momentum scale of GWs is necessarily related to the PBH mass, Eq.~\eqref{eq:pbhmass}.
Also, the amount of GWs is roughly proportional to the square of the scalar perturbation,
which may be conveniently estimated as
$\Omega_\text{GW} \sim 10^{-9} ({\mathcal P}_\zeta / 0.01)^2$.
The current pulsar timing array experiments~\cite{Arzoumanian:2015liz,Lentati:2015qwp,Shannon:2015ect} 
put severe constraints on $k \sim 10^{6}$\,Mpc$^{-1}$ corresponding to 
$M \sim 0.75 \gamma M_\odot$--$75 \gamma M_\odot$.
If one would like to interpret the LIGO events as PBH-mergers, these constraints play important roles~\cite{Inomata:2016rbd,Nakama:2016gzw,Orlofsky:2016vbd}.

%%%%%%%%%%%%%%%%%%%%%%%%%%%%%%%%%%%%%%%%%%%%
\section{PBH as All DM}
\label{sec:pbhasdm}
%%%%%%%%%%%%%%%%%%%%%%%%%%%%%%%%%%%%%%%%%%%%

\begin{figure}
	\centering
	\includegraphics[width=.45\textwidth]{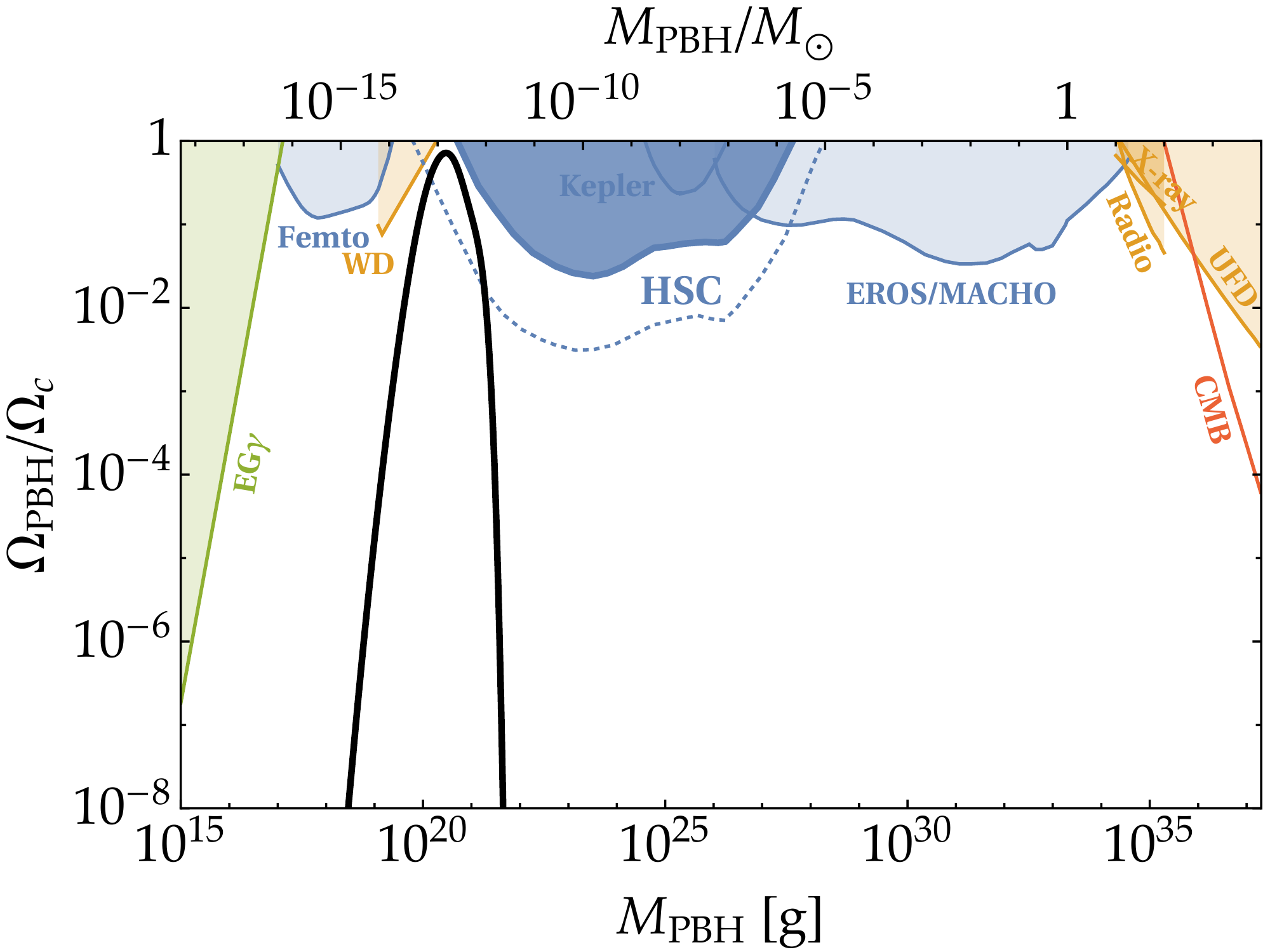}
	\caption{\small
	{\bf Black thick line}: $\Omega_\text{PBH} (M)$ for parameters given in Eq.~\eqref{eq:prms} is shown.
	We require the total abundance be equal to the observed DM density, $\Omega_\text{PBH,tot} = \Omega_c$.
	{\bf The solid lines with shades} represent relevant observational constraints on the current PBH mass spectrum [class (a)]:
	extra-galactic gamma-ray (EG$\gamma$)~\cite{Carr:2009jm}, 
	femtolensing (Femto)~\cite{Barnacka:2012bm}, 
	existence of white dwarfs in our local galaxy (WD)~\cite{Graham:2015apa},
	Subaru HSC microlensing (HSC)~\cite{takada}, 
	Kepler milli/microlensing (Kepler)~\cite{Griest:2013esa},  
	EROS/MACHO microlensing (EROS/MACHO)~\cite{Tisserand:2006zx},
	dynamical heating of ultra-faint dwarf galaxies (UFD)~\cite{Brandt:2016aco},
	and X-ray/radio constraints~\cite{Gaggero:2016dpq}.
	{\bf The solid line without shade} illustrates the observational constraints on the past PBH mass spectrum [class (b)]: 
	accretion constraints by CMB~\cite{Ali-Haimoud:2016mbv,Blum:2016cjs,Horowitz:2016lib}.
	Here we do not show the pulsar timing array constraints~\cite{Arzoumanian:2015liz,Lentati:2015qwp,Shannon:2015ect} 
	on gravitational waves via second order effects~\cite{Saito:2008jc,Saito:2009jt,Bugaev:2009zh,Bugaev:2010bb}
	because they are indirect and depend on the concrete shape of the scalar power spectrum.	Nevertheless, it is noticeable that their constraints are so strong that 
	PBHs with $M \sim 0.75 \gamma M_\odot$--$75 \gamma M_\odot$ are excluded
	(See for instance Fig.~1 in \cite{Inomata:2016rbd}),
	if they are generated via superhorizon fluctuations. See \cite{Inomata:2016rbd,Nakama:2016gzw,Orlofsky:2016vbd} for details.
	The conservative bound of the new HSC microlensing constraint is shown by the thick blue line with the deep shade,
	and the dotted one utilizes an extrapolation from the HST PHAT star catalogs in the disk region~\cite{takada}.
	}
	\label{fig:pbh}
\end{figure}

\begin{figure}
	\centering
	\includegraphics[width=.45\textwidth]{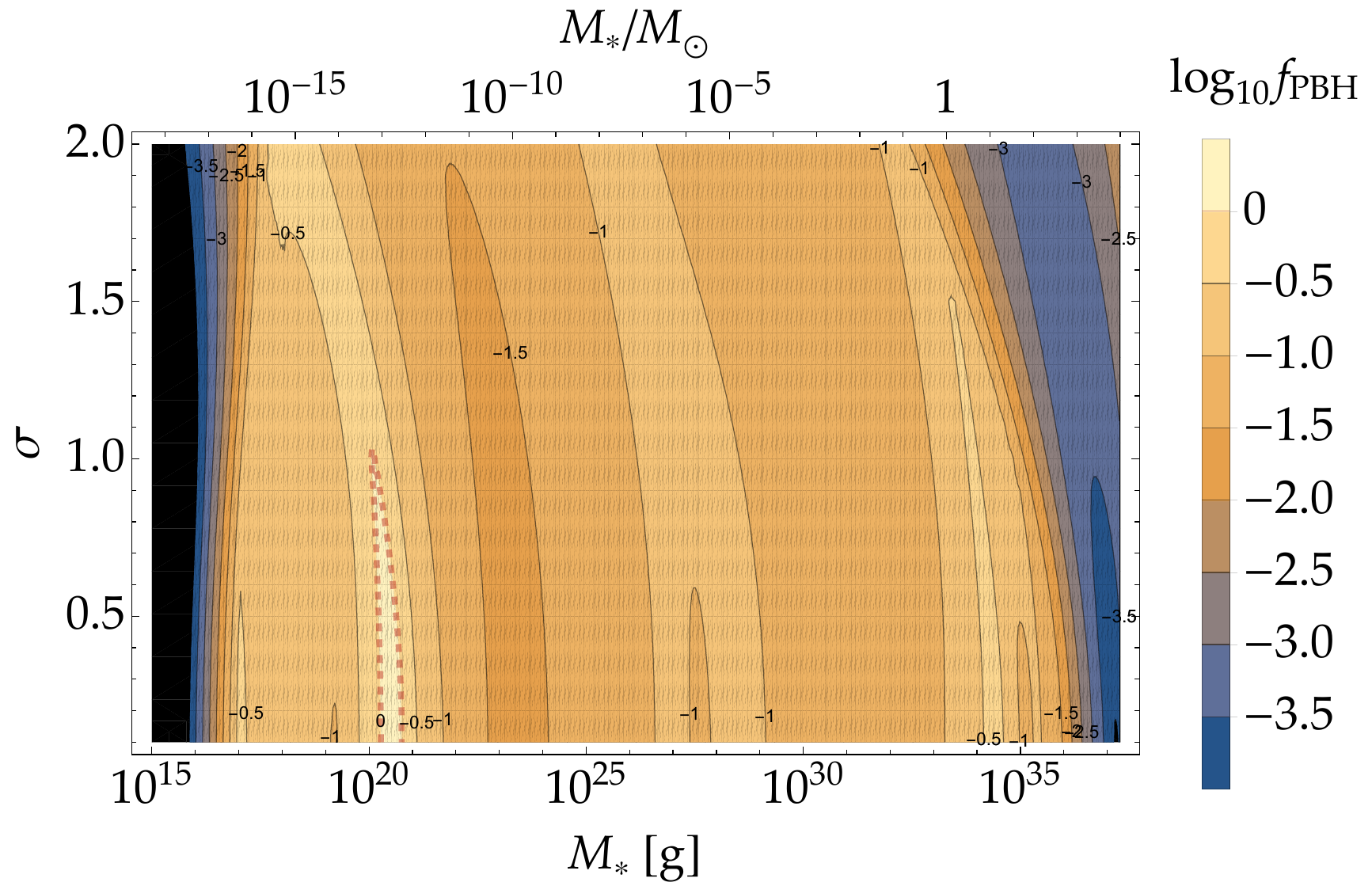}
	\caption{\small
	Constraints on parameters  $(M_\ast, \sigma, f_\text{PBH})$ of the extended mass function
	given in Eq.~\eqref{eq:extended}.
	Here we have adopted all the constraints shown in Fig.~\ref{fig:pbh}.
	The region consistent with the full DM, $f_{\text{PBH}} = 1$, is inside the dashed line
	near $M_\ast \simeq 10^{20}$\,g with
	$\sigma \lesssim 1$.
	}
	\label{fig:full_scan}
\end{figure}

As one can infer from Fig.~\ref{fig:pbh},
there are very limited ranges of the PBH mass in which PBHs can be a dominant component of DM.
The first viable region may lie between the white dwarf and HSC
 constraints around $\sim 10^{20}$\,\textrm{g}.\footnoteref{ns}
The next possibility would be between the MACHO/EROS and the
dynamical heating constraints around $10^{34-35}$\,\textrm{g}~\cite{Carr:2016drx},
since the CMB constraints can be much weaker as claimed recently~\cite{Ali-Haimoud:2016mbv,Blum:2016cjs,Horowitz:2016lib}.
This region is recently revisited because there is a possibility 
to explain the LIGO gravitational events simultaneously~\cite{Bird:2016dcv,Clesse:2016vqa,Sasaki:2016jop}.
However,
in Ref.~\cite{Green:2016xgy}, it is argued that PBHs as all DM in this region is disfavored
if one uses the constraint from the dynamical heating of ultra-faint dwarf galaxies.\footnote{
	Ref.~\cite{Carr:2016drx} varies the parameters of constraints from the dynamical heating of Eridanus II.
}
Ref.~\cite{Gaggero:2016dpq} also claims 
that PBHs cannot constitute all the DM for $M \sim \mathcal O (10) M_\odot$
by using a new accretion constraint on PBHs at the galactic center via the radio and X-ray.\footnote{
	Note that this constraint depends on the profile of PBH DM.
	For the Burkert profile, we can evade it as discussed in Ref.~\cite{Gaggero:2016dpq}.
}
In addition,
for PBHs generated via superhorizon fluctuations, 
the pulsar timing array experiments~\cite{Arzoumanian:2015liz,Lentati:2015qwp,Shannon:2015ect} 
set severe constraints  on gravitational waves via the second order effects~\cite{Saito:2008jc,Saito:2009jt,Bugaev:2009zh,Bugaev:2010bb}
for $M \sim 0.75 \gamma M_\odot$--$75 \gamma M_\odot$ as mentioned previously.
If the formation of PBHs is well approximated by the Gaussian statistics,
the power spectrum of curvature perturbation should be sharp enough to avoid the constraints at $\mathcal O(10) M_\odot$~\cite{Inomata:2016rbd,Orlofsky:2016vbd}.
Inflation models with enhanced non-Gaussianity at small scales
may evade this constraint 
since the same amount of PBHs can be produced by a smaller amplitude of the curvature perturbation than the Gaussian one~\cite{Nakama:2016gzw}.
We will return to these issues elsewhere~\cite{eleswhere}.

\begin{figure}
	\centering
	\includegraphics[width=.45\textwidth]{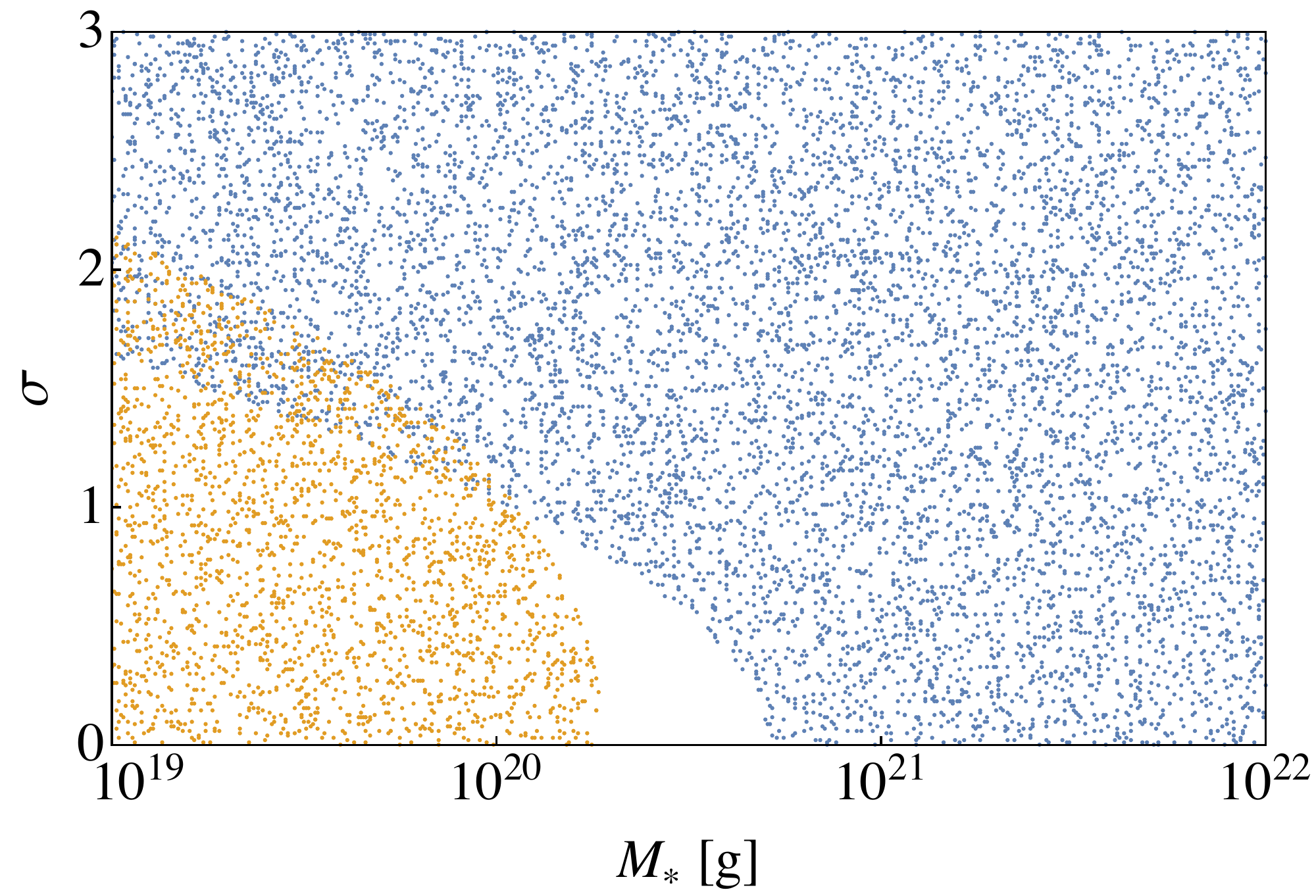}
	\caption{\small
	Constraints on parameters of the extended mass function given in Eq.~\eqref{eq:extended}
	with $M_\ast \simeq 10^{20}$\,g and $f_\text{PBH} = 1$.
	{\bf Orange dotted region} is excluded by WD~\cite{Graham:2015apa}.
	{\bf Blue dotted region} is excluded by HSC~\cite{takada}.
	}
	\label{fig:proto}
\end{figure}

Fig.~\ref{fig:full_scan} shows observational constraints shown in Fig.~\ref{fig:pbh}
on parameters of the following form of the extended mass function adopted in Ref.~\cite{Green:2016xgy}:
\begin{align}
	\frac{\mathrm d}{\mathrm d M}
	\frac{\Omega_\text{PBH} (M)}{\Omega_c}
	= N
	\exp \left[ \frac{-(\log M-\log M_*)^2}{2 \sigma^2} \right],
	\label{eq:extended}
\end{align}
where $N$ is determined so that the integration of Eq.~\eqref{eq:extended} becomes $f_\text{PBH}$.
This results indicate that, if we employ constraints shown in Fig.~\ref{fig:pbh},
the viable region is $M \sim 10^{20}$\,g.
Moreover, one can see that
the PBH mass spectrum should be sharp enough, $\sigma \lesssim 0.1$, to avoid the new constraint.
A closer view of this region is displayed in Fig.~\ref{fig:proto} for $f_\text{PBH} = 1$.
Such a sharp peak in the curvature perturbation can be obtained for a sizable $\kappa$.
See also discussion around Eq.~\eqref{eq:power_spec}.
We explicitly provide parameters of our model:
\begin{align}
	n&=3, \quad \frac{v}{\Mpl}=10^{-3}, \quad \kappa=0.13, \quad \alpha=\frac{\varepsilon\Mpl^2}{v^2}=0.640, \nonumber \\ 
	g&=5.44\times10^{-4}, \quad c_\text{pot}=1, %\quad c_\text{kin}=0,
	\label{eq:prms}
\end{align}
which satisfies $\Omega_\text{PBH,tot} = \Omega_c$.
We have checked that the predicted extended mass function passes all the observational constraints. See also Fig.~\ref{fig:pbh}.
One can see that our inflation model can produce
PBHs with $ \sim 10^{20}$\,g responsible for all DM,
if one takes the conservative bound given in Ref.~\cite{takada}.

%%%%%%%%%%%%%%%%%%%%%%%%%%%%%%%%%
%%%%%%%%%%% Conclusions  %%%%%%%%%%%
%%%%%%%%%%%%%%%%%%%%%%%%%%%%%%%%%
\section{Conclusions}
\label{sec:conc}

In this paper, we have revisited the idea of PBH as all DM,
confronting the new severe microlensing constraint in Ref.~\cite{takada}.
The new constraint puts the severest upper bounds on the abundance of PBHs from $10^{20}$\,g to $10^{28}$\,g.
We have shown that 
the remaining mass window for PBH as all DM is around $\simeq 10^{20}$\,g
and also that the width of the mass function should be sharp enough $\sigma \lesssim 0.1$,
by treating constraints on the extended mass function in a proper way.
We have constructed an inflationary model which yields a sharp PBH mass spectrum at $10^{20}$\,g and avoids the observational constraints.
Although it is quite marginal, it has been shown that the PBH can still be responsible for all DM.

Interestingly, a parameter set with $v \gtrsim 10^{15}$\,GeV is allowed,
which is compatible with a scenario in Ref.~\cite{Kawasaki:2016ijp}.
There, three of the authors proposed a mechanism,
where the Higgs metastability problem during inflation~\cite{Espinosa:2007qp,Fairbairn:2014zia,Hook:2014uia,Herranen:2014cua,Kamada:2014ufa,Kearney:2015vba,Espinosa:2015qea,East:2016anr} 
and preheating~\cite{Ema:2016kpf,Kohri:2016wof,Enqvist:2016mqj}
can be avoided and PBHs are simultaneously generated, while the SM sector is kept intact.
Therefore, a minimal scenario, 
involving the metastable electroweak vacuum,
chaotic inflation, and PBHs as all DM, is still viable.

Since the new constraint only utilizes one-night data observed by Subaru HSC~\cite{takada}, further observations might completely close the remaining window at $10^{20}$\,g.
As claimed in Ref.~\cite{Graham:2015apa},
once populations of relatively heavy white dwarfs at the galactic center are confirmed, constraints from the existence of white dwarfs
may also close the remaining window.
Also, as revisited in Ref.~\cite{Inomata:2016rbd} in the context of explaining LIGO events by PBH-mergers,
the induced GW with $f \sim \mathcal O (10^{-2})$\,Hz could be an interesting signature 
for future space-based detectors~\cite{Seoane:2013qna,Harry:2006fi,Seto:2001qf}.
PBH is one of leading non-particle candidates of DM.
Thus, for particle physics, it is of quite importance to probe/exclude the remaining window at $\simeq 10^{20}$\,g via several observations.

%%%%%%%%%%%%%%%%%%%%%%%%%%%%%%%%%%
%%%%%%%%%%% Acknowledge %%%%%%%%%%%
%%%%%%%%%%%%%%%%%%%%%%%%%%%%%%%%%%
\section*{Acknowledgements}
\small\noindent
This work is supported by Grant-in-Aid for Scientific Research from the Ministry of Education,
Science, Sports, and Culture (MEXT), Japan,  No.\ 15H05889 (M.K.), No.\ 25400248 (M.K.),
No.\ 26104009 (T.T.Y.), No.\ 26287039 (T.T.Y.) and No.\ 16H02176 (T.T.Y.), 
World Premier International Research Center Initiative (WPI Initiative), MEXT, Japan 
(K.I., M.K., K.M., Y.T., and T.T.Y.),
JSPS Research
Fellowships for Young Scientists (Y.T., and K.M.), and Advanced
Leading Graduate Course for Photon Science (K.I.).

\appendix
%%%%%%%%%%%%%%%%%%%%%%%%%%%%%%%%%%
%%%%%%%%%%% Appendix %%%%%%%%%%%
%%%%%%%%%%%%%%%%%%%%%%%%%%%%%%%%%%

\section{Constraints on extended mass functions}
As emphasized in the main text, though a physical mass function of PBHs is 
smooth and continuous, all the constraints summarized in Fig.~\ref{fig:pbh} 
assume that it takes a delta function. 
And thus, we need to be careful in applying these constraints on the extended mass function.
In this section, we revisit this issue for the sake of completeness.

Let us start with the following simplest example.
Suppose that we have a constant upper bound on 
the PBH fraction $\bar f_{i:i+1}$ between $M_i$ and $M_{i+1}$
assuming a monochromatic mass function.
The question is how to apply it to the extended mass function $f(M)$.
In this case, the constraint is translated into the following upper bound:
\begin{align}
	\bar N > N(f) \to
	1 > \frac{1}{\bar f_{i:i+1}} \int_{M_i}^{M_{i+1}} \mathrm d \ln M f(M).
\end{align}
Here $\bar N$ is an observational upper bound on the event number.
We assume that the expected event number $N(f)$ is expressed as an integration of
some function that is linear in $f$. 
%\TODO{KM: True?: This property is satisfied for the constraints summarized in Fig.~\ref{fig:pbh}.}
The next simplest example may be the following.
We have not only $\bar f_{i:i+1}$ in $[M_i, M_{i+1}]$ but $\bar f_{i-1:i}$ in $[M_{i-1}, M_{i}]$
which come from the same observation.
Then, the upper bound on the extended mass function would be
\begin{align}
%	\bar N > N(f) \to
	1 > \frac{1}{\bar f_{i-1:i}} \int_{M_{i-1}}^{M_{i}} \mathrm d \ln M f(M) +
	\frac{1}{\bar f_{i:i+1}} \int_{M_i}^{M_{i+1}} \mathrm d \ln M f(M).
\end{align}

The above consideration suggests the following procedure for a more generic upper bound
expressed as a continuous function.
First, pick up one observational constraint $x$, take small enough mass bins, 
and split the upper bound $\bar f^{(x)} (M)$ into a set of constant upper bounds
in each mass bin: $\{ \bar f_{i:i+1}^{(x)} ~\text{for}~ [M_i, M_{i+1}]; i = 1,\dots,N-1 \}$.
Then, the constraints of $x$ on the extended mass function $f(M)$ can be expressed as
\begin{align}
	1 > \sum_{i=1}^{N-1} \int_{M_i}^{M_{i+1}} \mathrm d \ln M \frac{f (M)}{f_{i:i+1}^{(x)}}
	\to 1 > \int_{M_1}^{M_N} \ln M \frac{f (M)}{f^{(x)} (M)}.
\end{align}
In the second equation, we assume small enough mass bins and approximate it
by a continuous integration from $M_1$ to $M_N$,
which gives a proper way to constrain PBHs with an extended mass function
from a given continuous upper bound of an observation $x$, $\bar f^{(x)} (M)$,
In the main text, we have used this second equation as a convenient way to constrain the extended mass function.
Finally, by imposing this bound for all the constraints (all $x$),
we can discuss the allowed region of the extended mass function
parametrized in Eq.~\eqref{eq:extended}.

%%%%%%%%%%%%%%%%%%%%%%%%%%%%%%%%%
%%%%%%%%%%% References %%%%%%%%%%%
%%%%%%%%%%%%%%%%%%%%%%%%%%%%%%%%%
\small
\bibliographystyle{apsrev4-1}
\bibliography{pbh_alldm}
\end{document}